# Azimuthally polarized terahertz radiation generation using radially polarized laser pulse in magnetized plasma

Shivani Aggarwal, Dinkar Mishra, Saumya Singh, Bhupesh Kumar[1], Pallavi Jha[2]

Department of Physics, University of Lucknow, Lucknow, Uttar Pradesh - 226007

[1]Corresponding Author: bhupeshk05@gmail.com

[2]Retired Professor

## Abstract

An analytical formulation of a radially polarized laser pulse propagating in a homogeneous, magnetized plasma is presented using Lorentz force, continuity and Maxwell's equations. Perturbation technique and quasi-static approximation (QSA) have been used to study the generated fields in nonlinear regime. The generated slow, oscillating, transverse electric and magnetic fields ($E_\theta$ and $B_r$) having equal amplitude, constitute a radiation field having frequency in the terahertz (THz) range. Particle-in-cell (PIC) simulation code FBPIC is used to validate analytical findings. Simulation studies also show that the generated THz radiation field propagates beyond the plasma boundary, indicating coherent electromagnetic radiation emission. Furthermore, the field amplitude scales nonlinearly with plasma density and increases linearly with external magnetic field strength, highlighting the role of these parameters in controlling radiation amplitude.

## 1. Introduction

The interaction of high-intensity laser pulses with homogeneous or inhomogeneous plasma gives rise to various nonlinear phenomena, forming the basis of numerous advanced applications such as laser-driven wakefield acceleration [1-2], inertial confinement fusion [3–4], and radiation generation having frequency ranging from x-ray to terahertz [5-8]. In a seminal work in 1979, Tajima and Dawson demonstrated that a high-intensity laser pulse propagating through plasma induces a ponderomotive force, which in turn drives the formation of electron-plasma waves known as wakefields [1]. In recent years, extensive efforts have been devoted to generating terahertz (THz) radiation using laser-plasma interaction in various configurations. Terahertz generation enables diverse applications in biomedical imaging [9], security screening [10], high-speed wireless communications [11], and material characterization [12].

Several analytical studies have shown THz radiation generation via propagation of laser pulses in plasma. Wolfinger et al. investigated the generation of axially polarized THz pulses via interaction of linearly polarized laser with plasma [13]. Nonlinear interaction of p-polarized chirped laser pulses in hot plasmas, reinforcing the idea that thermal and density gradients can be optimized to support frequency tunability in THz emission have been studied [14]. Peñano et al. investigated the role of relative polarizations and pulse duration of two-color laser pulses and collisional effects in plasma on THz generation [15]. Balakin et al. investigated THz emission from plasma driven by two-color femtosecond laser pulses, highlighting the roles of ponderomotive and radiation pressure forces [16]. Analytical study has demonstrated that radially polarized THz radiation can be emitted from laser-induced plasma wakes, where oscillating radial currents in plasma bubbles generate highly collimated THz beams. The emission properties of THz depend strongly on plasma thickness and laser intensity [8].

THz radiation generation by laser pulses propagating in magnetized plasma has also been reported. Several other studies on THz generation via laser-plasma interaction in various configurations have been reported [17-19]. Saroch et al. performed a simulation study on THz radiation generation by circularly polarized laser pulses propagating in axially magnetized plasma [20]. Tunable THz radiation generation in axially magnetized plasma channel has been reported using circularly polarized laser pulses [21]. Interaction of circularly polarized laser pulse with axially magnetized plasmas has shown that dominant radial field components of THz are generated under relativistic laser wakefield excitation [22]. Analytical study shows that arbitrarily polarized multi-color laser pulses propagating in obliquely magnetized plasma results in THz generation [7]. Mishra et al. demonstrated the generation of twisted THz radiation using a circularly polarized Laguerre–Gaussian laser pulse in axially magnetized plasma, revealing that the superposition of orthogonal THz components produces a linearly polarized twisted beam whose characteristics depend on the magnetic field and laser mode indices [23]. Further, simulation studies by Wu et. al. have shown that radially polarized laser pulses efficiently generate THz radiation by driving strong plasma currents [24].

In the context of plasma-based THz sources, radially polarized lasers may be of potential use due to their ability to induce strong longitudinal electric fields and cylindrically symmetric current distributions. This motivates study of radially polarized laser pulse interaction with magnetized plasma for THz generation. This work presents an investigation of terahertz radiation generation driven by radially polarized laser pulses in axially magnetized plasma, emphasizing the role of radial field symmetry in enhancing coherent THz emission. The paper is organized as follows: Section 2 presents the mathematical formulation describing the interaction between radially polarized laser pulses and axially magnetized plasma, hence establishing the theoretical foundation for THz radiation generation. The analytical framework for THz emission is developed, incorporating the role of radial polarization and plasma parameters. Section 3 includes simulation studies that validate the theoretical predictions and provide deeper insights into spatial evolution of the emitted THz radiation, inside and beyond the plasma boundary. Section 4 summarizes the key findings and discusses the implications of the results along with possible directions for future research.

## 2. Mathematical formulation for terahertz radiation generation

Consider a radially polarized laser pulse propagating through homogeneous plasma, embedded in a constant external axial magnetic field $(\vec{b} = \hat{z}b_o)$. The electric $(\vec{E})$ and magnetic $(\vec{B})$ fields of the laser pulse propagating along the z- direction can be written as

$$\vec{E} = E_0 exp\left(\frac{-r^2}{r_o^2}\right) g(z,t) \left(\hat{r}\left\{\left(\frac{r}{2r_0}\right) cos(k_o z - \omega_o t)\right\} + \hat{z}\left\{\left(1 - \frac{r^2}{r_o^2}\right) sin(k_o z - \omega_o t)\right\}\right), \qquad (1)$$

$$\vec{B} = \hat{\theta} E_0 \frac{r}{2r_0} exp\left(\frac{-r^2}{r_o^2}\right) g(z,t) cos(k_o z - \omega_o t), \qquad (2)$$

where $E_0$ is the amplitude of the laser field, $g(z,t) = exp(-(z-ct)^2/L^2)$ represents the Gaussian longitudinal envelope of the laser pulse and $c, L, k_o, r_o, \omega_o$ are respectively the speed of light in vacuum, pulse length, wave number, spot size and frequency of the radially polarized laser pulse. The motion of plasma electrons in the presence of these fields is governed by the Lorentz force and continuity equations

$$\frac{d\vec{v}}{dt} = \frac{\partial \vec{v}}{\partial t} + (\vec{v}.\vec{\nabla})\vec{v} = \frac{-e}{m}\left[\vec{E} + \frac{1}{c}(\vec{v} \times \vec{B})\right], \qquad (3)$$

and

$$\frac{\partial n}{\partial t} + \vec{\nabla}.(n\vec{v}) = 0, \qquad (4)$$

where $n$ and $\vec{v}$ represent the perturbed plasma density and electron velocity in plasma. Perturbation technique is used to expand various parameters in orders of the laser strength parameter $a_o \left(= \frac{eE_o}{mc\omega_o} \ll 1\right)$. The first order quiver velocity components and density of the plasma electrons are obtained with the help of Eqs. (1) and (2) as,

$$v_r^{(1)} = \frac{-a_o c}{(1-\Omega^2)} A_1 \sin(k_o z - \omega_o t), \tag{5}$$

$$v_\theta^{(1)} = \frac{-a_o c \Omega}{(1-\Omega^2)} A_1 \cos(k_o z - \omega_o t), \tag{6}$$

$$v_z^{(1)} = -\left(1 + \frac{\omega_p^2}{\omega_o^2}\right) a_o c A_2 \cos(k_o z - \omega_o t), \tag{7}$$

$$n^{(1)} = -n_o a_o A_2 \cos(k_o z - \omega_o t), \tag{8}$$

where $A_1 = \frac{r}{2r_0} \exp\left(\frac{-r^2}{r_0^2}\right) g(z,t)$, $A_2 = \left(1 - \frac{r^2}{r_0^2}\right) \exp\left(\frac{-r^2}{r_0^2}\right) g(z,t)$, $\Omega (= \omega_c/\omega_o)$ is the normalized cyclotron frequency and $n_o$ is the ambient plasma density. It may be noted that in the absence of the magnetic field, quiver velocity components and density reduce to the standard quiver velocity and density of the plasma electrons in unmagnetized plasma [25].

The first order, fast oscillating perturbations of plasma electron velocity and density lead to second order slow oscillation of these parameters. Eq. (3) is used to define the evolution of the second order velocities governed by

$$\frac{\partial v_z^{(2)}}{\partial t} = -\frac{e}{m} E_z^{(2)} - \frac{1}{2} \frac{\partial}{\partial z}\left(\vec{v}^{(1)} \cdot \vec{v}^{(1)}\right), \tag{9a}$$

$$\frac{\partial v_r^{(2)}}{\partial t} = -\frac{e}{m} E_r^{(2)} + \frac{e}{mc}\left(v_z^{(1)} B_\theta^{(1)}\right) - \frac{e}{mc}\left(v_\theta^{(2)} b_z\right) - \frac{1}{2} \frac{\partial}{\partial r}\left(\vec{v}^{(1)} \cdot \vec{v}^{(1)}\right) + \left[\vec{v}^{(1)} \times \left(\vec{\nabla} \times \vec{v}^{(1)}\right)\right]_r,$$

(9b)

$$\frac{\partial v_\theta^{(2)}}{\partial t} = -\frac{e}{m} E_\theta^{(2)} - \frac{e}{mc}\left(v_r^{(2)} b_z\right) + \left[\vec{v}^{(1)} \times \left(\vec{\nabla} \times \vec{v}^{(1)}\right)\right]_\theta, \tag{9c}$$

where $E^{(2)}_{r,\theta,z}$ represents the generated slow, transverse and longitudinal electric wakefields. It may be noted that the nonlinear ponderomotive term $[\vec{\nabla}(\vec{v}.\vec{v})]$ contributes to transverse $(E^{(2)}_r)$ as well as longitudinal $(E^{(2)}_z)$ wakefield generation. In addition, coupling of the external magnetic field with transverse velocity components drives the slow radial $(E^{(2)}_r)$ and azimuthal $(E^{(2)}_\theta)$ electric fields.

In order to evaluate the slowly evolving wakefields, Maxwell's equations are set up in the transformed frame ($\xi = z - ct, \tau = t$) and Eq. (9) is also transformed to the same reference frame. Evolution of wakefields can be studied in the laser pulse frame using above transformation of coordinates. Further quasi-static approximation (QSA) is applied to the Maxwell's equations. Under QSA plasma electron experiences nearly a static field [26]. Thus

$$\frac{\partial E_\theta}{\partial \xi} = \frac{-\partial B_r}{\partial \xi}, \tag{10a}$$

$$\frac{\partial E_r}{\partial \xi} - \frac{\partial E_z}{\partial r} = \frac{\partial B_\theta}{\partial \xi}, \tag{10b}$$

$$\frac{1}{r}\left[\frac{\partial (rE_\theta)}{\partial r}\right] = \frac{\partial B_z}{\partial \xi}, \tag{10c}$$

and

$$\frac{-\partial B_\theta}{\partial \xi} = \frac{4\pi}{c}J_r - \frac{\partial E_r}{\partial \xi}, \tag{11a}$$

$$\frac{1}{r}\frac{\partial (rB_\theta)}{\partial r} = \frac{4\pi}{c}J_z - \frac{\partial E_z}{\partial \xi}, \tag{11b}$$

$$\frac{\partial B_r}{\partial \xi} - \frac{\partial B_z}{\partial r} = \frac{4\pi}{c}J_\theta - \frac{\partial E_\theta}{\partial \xi}, \tag{11c}$$

where $J^{(2)}_{r,\theta,z} \left(= \left(-en_o v^{(2)}_{r,\theta,z}\right)\right)$ are the transverse and longitudinal current densities respectively. While deriving Maxwell's equations in the transformed co-ordinates, the generated fields are assumed to be axisymmetric $\left(\frac{\partial}{\partial \theta} = 0\right)$. In order to obtain the current densities, the second order axial and radial components of velocity are derived using Eqs. (5), (6), (7), and (8).

Assuming that the laser spot size is larger than the laser pulse length. Under broad beam $(r_o > L)$ condition the transverse derivatives of transverse fields are neglected as compared to longitudinal derivatives. Thus Eqs. (10c) and (11b) reduce to

$$B_z^{(2)} = 0 \tag{12}$$

and

$$\frac{\partial}{\partial \xi} E_z^{(2)} = \frac{4\pi}{c} J_z^{(2)} \tag{13}$$

Differentiating equation (13) with respect to $\xi$ and combining with equation (9a) gives

$$\left(\frac{\partial^2}{\partial \xi^2} + k_p^2\right) E_z^{(2)} = \frac{4\pi n_o e}{c} P \xi \exp\left(-\frac{2\xi^2}{L^2}\right) \tag{14}$$

where $P = \frac{a_o^2 c}{L^2} \exp\left(\frac{-2r^2}{r_o^2}\right) \left[\left(\frac{r}{2r_o}\right)^2 \left(\frac{1+\Omega^2}{1-\Omega^2}\right) + \left(1 - \frac{r^2}{r_o^2}\right)^2\right]$, solving equation (14) gives the longitudinal electric wakefield generated behind ($\xi < 0$) the laser pulse, as

$$E_z^{(2)} = \frac{\omega_p^2 m}{e} \frac{La_o^2}{4} \sqrt{\frac{\pi}{2}} \exp\left(\frac{-2r^2}{r_o^2}\right) \exp\left(-\frac{k_p^2 L^2}{8}\right) \left[\left(\frac{r}{2r_o}\right)^2 \left(\frac{1+\Omega^2}{1-\Omega^2}\right) + \left(1 - \frac{r^2}{r_o^2}\right)^2\right] \cos(k_p \xi) \tag{15}$$

It may be noted that in the absence of external magnetic field, the electric wakefield reduces to that generated by a radially polarized laser pulse propagating in unmagnetized homogeneous plasma [25].

In order to obtain the transverse electric wakefields generated in magnetized plasma, Maxwell's equations (10b) and (11a) combined with (9c) are used alongwith Eq. (15) to give

$$E_r^{(2)} = \frac{\omega_p^2 m}{e} \frac{La_0^2}{4} \sqrt{\frac{\pi}{2}} \exp\left(\frac{-2r^2}{r_0^2}\right) \exp\left(-\frac{k_p^2 L^2}{8}\right) \left[\left(\frac{1+\Omega^2}{1-\Omega^2}\right)\left(\frac{r}{2r_0^2} - \frac{r^3}{r_0^4}\right) + \left(\frac{-4r}{r_0^2}\right)\left(1 - \frac{r^2}{r_0^2}\right)\left(2 - \frac{r^2}{r_0^2}\right)\right] \sin(k_p \xi) \qquad (16)$$

and Eqs. (11c) and (10b) combined with Eq. (9c) are used alongwith Eq. (15) to give

$$E_\theta^{(2)} = \frac{\Omega \omega_o}{ck_p^2} \frac{\omega_p^2 m}{e} \frac{La_0^2}{4} \sqrt{\frac{\pi}{2}} \exp\left(\frac{-2r^2}{r_0^2}\right) \exp\left(-\frac{k_p^2 L^2}{8}\right) \left[\left(\frac{1+\Omega^2}{1-\Omega^2}\right)\left(\frac{r}{2r_0^2} - \frac{r^3}{r_0^4}\right) + \left(\frac{-4r}{r_0^2}\right)\left(1 - \frac{r^2}{r_0^2}\right)\left(2 - \frac{r^2}{r_0^2}\right)\right] \cos(k_p \xi) \qquad (17)$$

Equation (15), (16) and (17) respectively represent the longitudinal ($E_z^{(2)}$), transverse ($E_r^{(2)}$), and azimuthal ($E_\theta^{(2)}$) wakefields generated behind the pulse. Now using Eq. (10), transverse component of generated magnetic field behind the pulse is given by,

$$B_r^{(2)} = -\frac{\Omega \omega_o}{ck_p^2} \frac{\omega_p^2 m}{e} \frac{La_0^2}{4} \sqrt{\frac{\pi}{2}} \exp\left(\frac{-2r^2}{r_0^2}\right) \exp\left(-\frac{k_p^2 L^2}{8}\right) \left[\left(\frac{1+\Omega^2}{1-\Omega^2}\right)\left(\frac{r}{2r_0^2} - \frac{r^3}{r_0^4}\right) + \left(\frac{-4r}{r_0^2}\right)\left(1 - \frac{r^2}{r_0^2}\right)\left(2 - \frac{r^2}{r_0^2}\right)\right] \cos(k_p \xi). \qquad (18)$$

Considering Eqs. (17) and (18), it may be noted that mutually perpendicular components of electric ($E_\theta^{(2)}$) and magnetic ($B_r^{(2)}$) fields oscillating at the plasma frequency and having equal amplitude, are obtained. If the plasma frequency lies in the THz range, $E_\theta^{(2)}$ and $B_r^{(2)}$ constituted a THz radiation field. It may be noted that the generated THz radiation is azimuthally polarized and its amplitude varies with the radial position. These field components vanish in the absence of the external magnetic field.

## 3. Simulation study and graphical analysis

In the present study, we perform fully electromagnetic, quasi-3D Fourier-Bessel particle-in-cell (FBPIC) simulations to investigate the nonlinear interaction of a radially polarized laser pulse with magnetized plasma. The simulation is performed using moving window in cylindrical geometry. The computational domain spans from z = −10 μm to 120 μm axially and up to r = 20 μm radially, resolved with 1200 and 150 grid points, respectively. The temporal resolution is determined by the Courant condition, ensuring numerical stability. These settings correspond to a spatial resolution that provides more than 10 grid points per laser wavelength (λ = 0.8 μm), allowing accurate representation of both the laser fields and the plasma response. A radially polarized laser pulse with normalized peak amplitude $a_o$ = 0.3, waist $r_o$ = 15 μm, and pulse duration τ = 40 fs is initialized at $z_0$ = 100 μm and launched into a homogeneous, fully ionized plasma extending from z = 120 μm to 250 μm and vacuum thereafter. The plasma has electron density of $n_o$ = 3.8×$10^{23}$ $m^{-3}$, with a linear density ramp over 20 μm at the entrance. Each grid cell is populated with 16 macroparticles (2 × 2 × 4 in z, r, and θ respectively) to ensure good statistical sampling of the distribution function. A static magnetic field of $b_o$ = 71 T is applied along the z-axis to model magnetized laser-plasma interaction. The combined spatial, temporal, and particle resolution enables accurate modeling of field generation as well as fine-scale electron dynamics.

Fig. 1 shows the radial variation of the azimuthal electric wakefield ($E_\theta$) with respect to $r$ obtained via analytical (dashed curve) and simulation (solid curve). The azimuthal electric field starts from zero at the axis ($r$ = 0), rises sharply with increasing $r$, and attains a maximum value of ≈ $8 \times 10^8 V/m$ at $r$ ≈ 5 μm, whereas the simulation peaks at a much lower magnitude of ≈ $0.9 \times 10^8 V/m$ at r ≈ 7 μm. The discrepancy in analytical and simulation results may be attributed to perturbative approximations and QSA. Further, simulation studies take into account of realistic laser pulse evolution alongwith energy depletion.

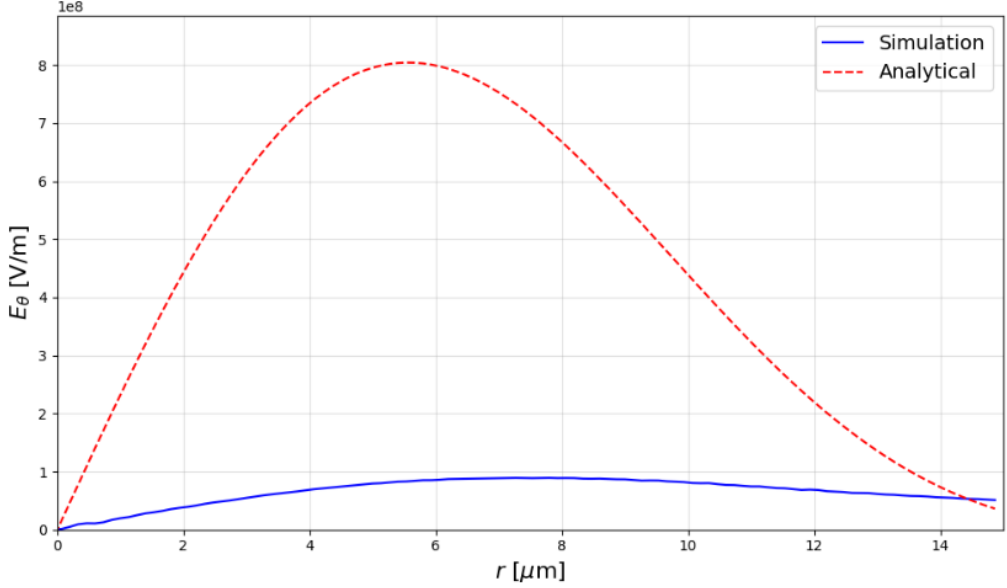

FIG. 1. Variation of amplitude of azimuthal electric field with respect to r for $a_o = 0.3$, $r_o = 15$ μm, L= 12 μm, $n_o = 3.8 \times 10^{23}$ m$^{-3}$, $b_o = 71\ T$, and $z = 250$ μm.

Figure 2 compares the analytically calculated azimuthal electric field E$_\theta$ (dashed) with simulation (solid) results at various radial position (r = 2 to 15 μm) along the propagation axis. The field amplitude varies significantly with r for analytical as well as simulation results. It is important to note that the THz radiation field propagates in vacuum (z > 250 μm) as seen via simulation results. However, analytical fields are truncated at the plasma-vacuum boundary. Hence simulation results present a realistic picture indicating the generation of a radiative field. By analyzing the periodicity in the simulation and analytical curves, the dominant spatial wavelength is approximately 55.4 μm, corresponding to a frequency (≈5.42 THz). The difference in analytical and simulation results obtained for generated THz may be attributed to the approximations considered in the analytical formulation.

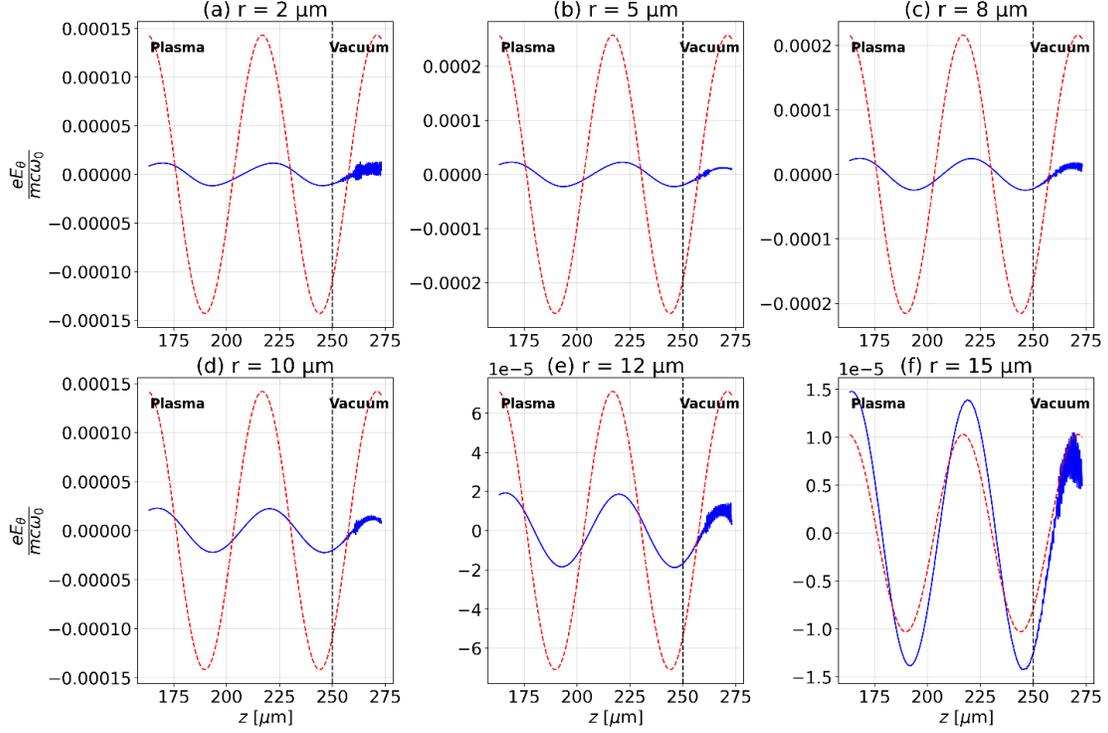

FIG. 2 Comparison of analytically calculated (dashed) and simulated (solid) normalized azimuthal electric field ($E_\theta$) at various radial positions (r) with plasma-vacuum boundary at 250 μm.

Figure 3 illustrates the radial magnetic field ($B_r$) at various radial position (r), comparing analytical results (dashed) with simulation results (solid). This field component, perpendicular to the propagation axis and orthogonal to the azimuthal electric field ($E_\theta$) shown in Fig. 2, demonstrates the generation of electromagnetic (EM) fields due to the interaction of a radially polarized laser pulse with the magnetized plasma. Notably, the fields persist beyond the plasma boundary at z = 250 μm and propagate in vacuum, further supporting the generation of terahertz radiation.

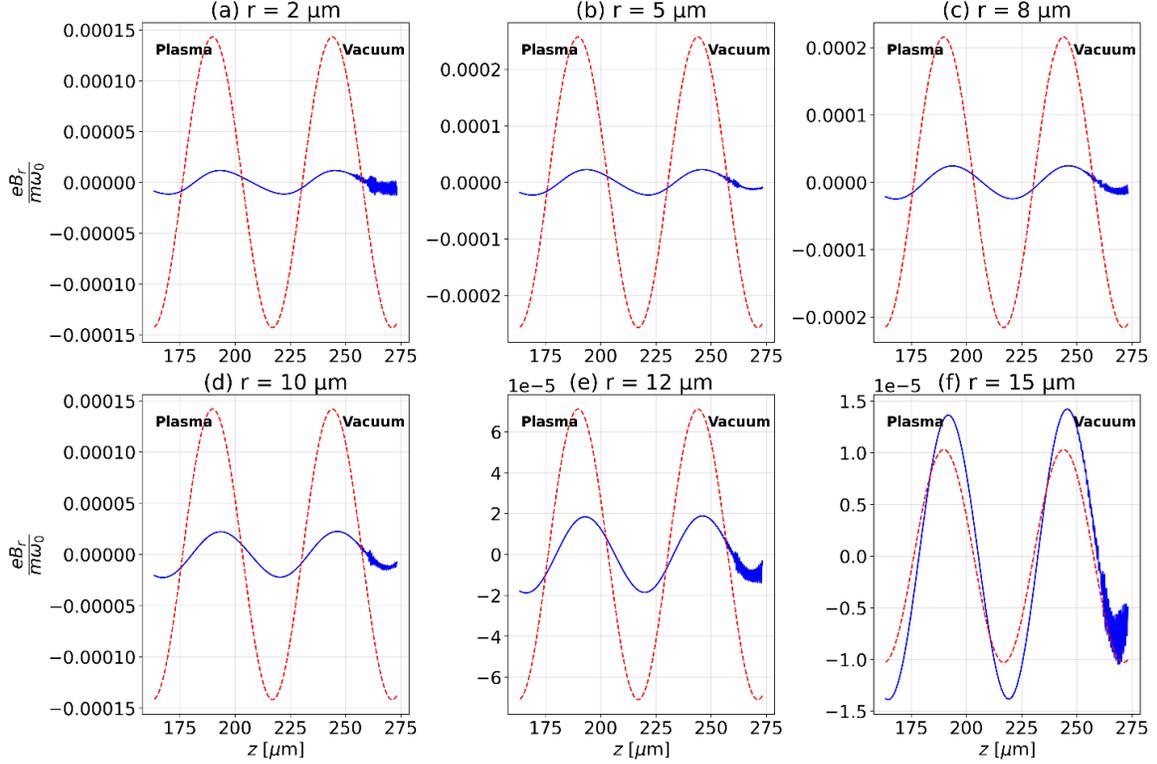

FIG. 3 Comparison of analytical (dashed) and simulation (solid) normalized radial magnetic field ($B_r$) at various radial position (r).

Figure 4 shows three-dimensional plot of the THz azimuthal electric field $E_\theta$ obtained from (a) PIC simulation and (b) analytical model using a radially polarized laser with $a_0$=0.3 and λ=0.8 μm interacting with a magnetized plasma ($n_0$=3.8×10$^{23}$ m$^{-3}$, $b_o$ =71 T). Simulation results show localized azimuthal field structures generated during laser–plasma interaction, while the analytical solution reproduces the key spatial features and oscillatory behavior of $E_\theta$. The close qualitative agreement validates the theoretical description of azimuthal field generation in magnetized plasma.

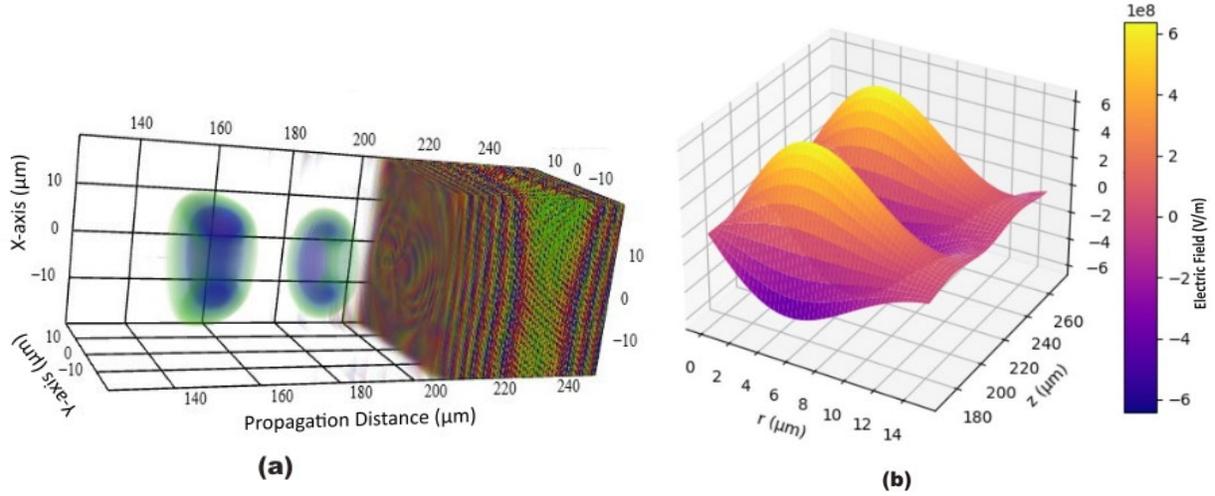

FIG. 4. 3D simulation (a) and analytical (b) plot of $E_\theta$ for $a_0 = 0.3$, $\lambda = 0.8$ μm, $n_o = 3.8\times10^{23}$m$^{-3}$, $b_o = 71$ T.

Figure 5 presents the simulated phase space profiles of the azimuthal electric field $E_\theta$ (Fig. 5(a)) and the radial magnetic field $B_r$ (Fig. 5(b)) during the interaction of a radially polarized laser pulse with magnetized plasma. The $E_\theta$ field exhibits periodic axial oscillations and strong localization near the beam axis, reflecting the coupling of radial polarization to azimuthal motion of electron. The $B_r$ distribution shows corresponding radial magnetic structures, generated by induced azimuthal currents in the presence of external magnetic field.

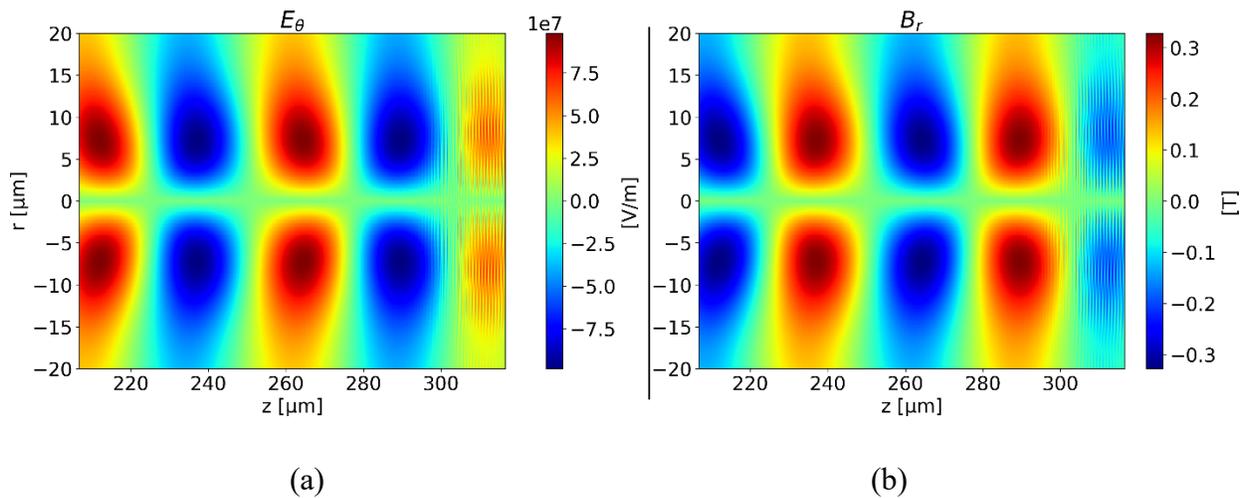

(a)          (b)

FIG. 5 Simulated profiles of $E_\theta$ and $B_r$ for a radially polarized laser ($a_0$ = 0.3, $\lambda$ = 0.8 μm) interacting with magnetized plasma ($n_o$ = 3.8×10$^{23}$m$^{-3}$, $b_o$ = 71 T). The periodic field structures indicate azimuthal current generation localized near the beam axis.

## 4. Summary and conclusion

This study analyses THz radiation generation using both analytical and FBPIC simulation methods. In this study interaction of a radially polarized laser pulse with a homogeneous, magnetized plasma has been considered. THz radiation fields are derived analytically by combining time-dependent Maxwell's equations, with slow, nonlinear plasma electron velocities coupled to the external magnetic field. Using Lorentz force and continuity equations, current density sources driving the fields are obtained. Applying QSA in the pulse frame yields both longitudinal and transverse wakefields. The azimuthal electric field ($E_\theta$) and magnetic field ($B_r$) are mutually orthogonal and equal in amplitude. Hence these transverse electric and magnetic field components oscillating at the plasma frequency behind the laser pulse leads to generation of azimuthally polarized THz radiation. The fields ($E_\theta$, $B_r$) exhibit cylindrical symmetry, vanish at the axis and vary with the radial position maximum at r ≈ 5 μm. Simulation studies show that the THz fields propagate beyond the plasma boundary (in vacuum). The discrepancy between analytical and simulation results are attributed to approximations considered in analytical formulation. The analytical prediction of THz radiation field generation is validated via simulation. The present study is significant since azimuthally polarized radiation has potential application in laser material processing [27], high-resolution microscopy [28], atomic lenses [29] and atom traps [30].